\documentclass[aps,prc,twocolumn,groupedaddress,showpacs]{revtex4}
\usepackage{graphicx}
\usepackage[rightcaption]{sidecap}

\begin{document}
\title{Refractive effects and Airy structure in  inelastic  $^{16}$O+$^{12}$C rainbow  scattering \\
}

\author{S. Ohkubo$^{1,2}$,  Y. Hirabayashi$^3$, A. A. Ogloblin$^4$,   
Yu. A. Gloukhov$^4$, \\
A. S. Dem'yanova$^4$ and
W. H. Trzaska$^5$ }
\affiliation{$^1$Research Center for Nuclear Physics, Osaka University, 
Ibaraki, Osaka 567-0047, Japan }
\affiliation{$^2$University of Kochi,  Kochi 780-8515, Japan  }
\affiliation{$^3$Information Initiative Center, Hokkaido University, Sapporo 060-0811, Japan}
\affiliation{$^4$RSC  ``Kurchatov Institute'',  RU-123182 Moscow, Russia}
\affiliation{$^5$ JYFL, FIN-40351 Jyv\"{a}skyl\"{a}, Finland}

\date{\today}

\begin{abstract}
\par
Inelastic $^{16}$O +$^{12}$C rainbow  scattering to the  $2^+$  (4.44 MeV)  state of 
$^{12}$C  was measured at the  incident energies, $E_L$ = 170, 181, 200, 260 and 281 MeV.  A systematic  analysis
 of the  experimental angular distributions was performed   using 
 the coupled channels method with an extended double folding  potential  
derived from  realistic  wave functions for $^{12}$C  and $^{16}$O calculated with 
a microscopic  $\alpha$ cluster model and   a finite-range  density-dependent 
  nucleon-nucleon force.
The coupled channels  analysis of the measured inelastic  scattering data
 shows consistently some Airy-like
structure in the inelastic scattering cross sections for the first $2^+$
state of $^{12}$C, which is somewhat obscured and still not clearly visible
in the measured data. 
The Airy minimum was identified from the analysis and the systematic
  energy  evolution of the  Airy structure   was studied. 
 The Airy minimum in inelastic scattering is found
to be shifted  backward compared with that in elastic scattering.
 \end{abstract}

\maketitle

\par
\section{INTRODUCTION}
Since the first observation of a nuclear rainbow in elastic $\alpha$ particle scattering from 
$^{58}$Ni \cite{Goldberg1974}, the importance of the   concept of rainbow scattering in studies 
of nuclear reactions and  structures
 has been widely understood \cite{Khoa2007,Michel1998,Brandan1997}.
 The interaction potential can be determined without discrete ambiguity
 up to the  internal region by studying  rainbow scattering because  refraction
  carries  information from the  inner region. The nuclear rainbow  
  has been observed under weak absorption in many systems  and has been extensively studied 
\cite{Khoa2007}.  The uniquely determined interaction potential    made it possible to
study the cluster structure of the compound system  above and below the threshold
 energy by unifying unbound and bound states as typically shown for the $\alpha$+$^{16}$O 
 and $\alpha$+$^{40}$Ca systems \cite{Michel1998}.
   The nuclear rainbow has also been  observed for  typical  heavy ion systems such as 
 $^{16}$O+$^{16}$O, $^{16}$O+$^{12}$C   and  $^{12}$C+$^{12}$C,  
for which  the  higher order Airy structure  has been clearly observed 
 \cite{Brandan1997,Ogloblin1998,Nicoli1999,Nicoli2000,Ogloblin2000,Szilner2001,Ogloblin2003}.
The interaction potential  
uniquely determined for the most  typical $^{16}$O+$^{16}$O system  made it possible
to understand the superdeformed $^{16}$O+$^{16}$O  cluster structure in $^{32}$S 
and the nuclear rainbow in a unified way \cite{Ohkubo2002,Ohkubo2003}.

\par
The nuclear rainbow has also been  observed for   inelastic scattering 
\cite{Bohlen1982,Bohlen1985,Brandan1986,Bohlen1993,Khoa2005,Khoa2007}.
 The  rainbow in   inelastic
 rainbow scattering makes it possible to  understand the interaction potential
 for the inelastic channels up to the internal region.
For  typical $\alpha$+$^{40}$Ca and
 $^6$Li+$^{12}$C  systems \cite{Michel2004A,Michel2007},  the mechanism of the nuclear rainbow and   the  Airy structure
 in inelastic scattering  has been  studied \cite{Michel2004A,Michel2005}.
 Inelastic rainbow scattering has also  been especially powerful in understanding the highly
 excited  cluster structure  near and above the threshold energy,  such as the $\alpha$ particle
 condensation of  the Hoyle state in $^{12}$C and the Hoyle-analog state in $^{16}$O
 \cite{Ohkubo2004,Ohkubo2007,Ohkubo2010,Belyaeva2010,Hamada2013}.
A recent systematic study of the evolution of the Airy structure in inelastic scattering 
 for the typical $\alpha$+$^{16}$O system \cite{Hirabayashi2013}     showed 
that  the cluster structure with core excitation in $^{20}$Ne near the threshold energy region 
and the inelastic nuclear  rainbow  scattering can be understood in a unified way by  using   
reliable  interaction potentials for the inelastic channels. 
This urges us to study inelastic rainbow scattering  with heavy ions  in order to determine
 the  interaction potentials in inelastic channels, which will  make it possible  to understand 
 the molecular 
 structure with core excitation, for which  phenomenological shallow potentials have been used widely
 instead of  a deep potential \cite{Abe1980}.

\par
For heavy ion systems, an  inelastic nuclear rainbow  has been observed
 for  $^{16}$O+$^{16}$O,  $^{16}$O+$^{12}$C  and  $^{12}$C+$^{12}$C scattering 
\cite{Bohlen1982,Bohlen1985,Brandan1986,Bohlen1993,Khoa2005,Khoa2007}.
For the typical $^{16}$O+$^{16}$O system, Khoa {\it et al.} \cite{Khoa2005}
 could  trace a weak rainbow pattern in the energy range  from 350-704 MeV,
 however, a   clear identification of  the Airy minimum and its energy evolution  
was not possible.
   For the asymmetric $^{16}$O+$^{12}$C system,  
  elastic scattering angular distributions  have been measured  without being  obscured due  to 
 symmetrization over a wide range of incident energies at $E_L=62-$1503 MeV
 \cite{Brandan1986,Ogloblin1998,Nicoli2000,Ogloblin2000,Szilner2001,Trzaska2002,Ogloblin2003}. 
Very recently, evidence for  a secondary
 bow in elastic scattering caused by coupling to the inelastic channel    has been reported 
 \cite{Ohkubo2014}. Therefore  it is  particularly intriguing  to study   inelastic rainbow
 scattering for this system.

\par
The purpose of this paper is to  study
  the existence and  evolution of the Airy structure
 in inelastic rainbow  scattering 
by  analyzing   the measured inelastic and elastic angular distributions of differential cross sections 
with an extended double folding model by using  realistic wave functions for $^{12}$C  and $^{16}$O
   obtained in the microscopic cluster model calculations.

\section{EXPERIMENT}
\par
  Differential cross-sections of  inelastic  $^{16}$O+$^{12}$C scattering at $E_L$= 170-280 MeV leading 
to the  $2^+$ (4.44 MeV)  state of $^{12}$C  were measured at the cyclotron of  Jyvaskyla University
 (Finland).  $^{16}$O beams with an intensity of about 100 nA (electrical) were exploited. 
The energetic resolution of the beam was $\approx$0.3\%, and the size of the beam spot on the target
 was about 2 mm$\times $3 mm. The targets were self-supporting carbon foils of  0.3 mg/cm$^2$
 thickness. 
A scattering chamber of  diameter $\approx$1500 mm was used in the experiment. Detectors were
 located on rotating tables, so the full angular range could be covered. 
    Differential cross-sections at  forward angles ($\theta_{c.m.}$= 7-40$^\circ$) were measured by
 a $\Delta E-E$ 
telescope of semiconductor counters. The thickness of $E$ and $\Delta E$ counters was
 600-800 and 20-40 $\mu$m, respectively. The solid angle covered by the telescope was about 
0.08 millisteradian. The total energy resolution (determined mainly by kinematics) was 1.2 MeV. 
The angular resolution was $\pm$0.2$^\circ$ and determined mainly by the beam angular 
spread.
      For  measurements at  larger angles with $\theta_{c.m.}$$>$$40^\circ$,  a 
position-sensitive $\Delta$$E-E$ detector was used. It included a gas-filled proportional $\Delta$$E$ counter
 with variable pressure, and an $E$ detector which consists of ten silicon pin-diodes of 
10 mm $\times$ 10 mm
 dimension each with thickness of 370-760$\mu$m. This detector could cover an angular range of
 about 10$^\circ$ in the laboratory system.
   The accuracy of the absolute cross-section measurements was estimated to be around 15\%. 

\begin{figure*}[thb]
\includegraphics[keepaspectratio,width=14cm] {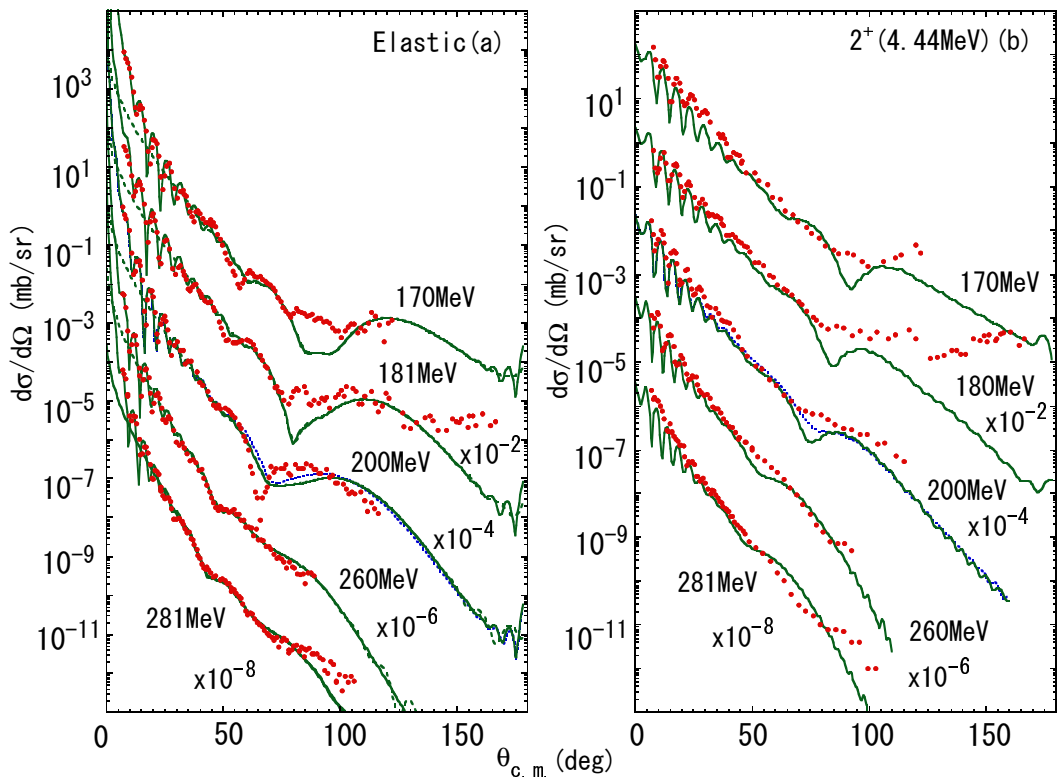}% angle=-90,,Here is how to import EPS art
 \protect\caption{\label{fig.1} {(Color online) 
Angular distributions in $^{16}$O+$^{12}$C   (a)  elastic and  (b) inelastic scattering 
to the $2^+$ state  of
 $^{12}$C  calculated using the 
coupled channels method  including coupling to the $2^+$ 
and $3^-$ states of $^{12}$C and $^{16}$O with  the imaginary potentials 
 in Table I (solid lines) are compared with the experimental data (points). 
 The green dashed lines in elastic scattering display the calculated farside components. 
The blue dotted lines at 200 MeV are calculated with  a channel dependent imaginary
potential for the $2^+$ and $3^-$ states of $^{12}$C  (see text).
 The elastic scattering data
are from Refs. \cite{Ogloblin2000,Ogloblin2003}.
 }
}
\end{figure*}

\section{ COUPLED CHANNELS ANALYSIS}
\par
We study    $^{16}$O+$^{12}$C  scattering  with the coupled channels   method  using 
an extended  double folding (EDF)  model that describes all the diagonal and off-diagonal
coupling potentials derived from  the microscopic   realistic wave functions for $^{12}$C  
and $^{16}$O  using  a density-dependent   nucleon-nucleon force.
  The diagonal and coupling potentials for the $^{16}$O+$^{12}$C system are calculated using
 the EDF  model  without introducing a normalization factor:
\begin{eqnarray}
\lefteqn{V_{ij,kl}({\bf R}) =
\int \rho_{ij}^{\rm (^{16}O)} ({\bf r}_{1})\;
     \rho_{kl}^{\rm (^{12}C)} ({\bf r}_{2})} \nonumber\\
&& \times v_{\it NN} (E,\rho,{\bf r}_{1} + {\bf R} - {\bf r}_{2})\;
{\it d}{\bf r}_{1} {\it d}{\bf r}_{2} ,
\end{eqnarray}
\noindent where $\rho_{ij}^{\rm (^{16}O)} ({\bf r})$ is the diagonal ($i=j$)  or transition ($i\neq j$)
 nucleon  density of  $^{16}$O    taken from  the microscopic $\alpha$+$^{12}$C  cluster model 
 wave functions calculated  with  the orthogonality 
condition model (OCM) in Ref. \cite{Okabe1995}. This model uses  a  realistic size parameter both 
 for the $\alpha$ particle and $^{12}$C  and is an extended version of  the  
OCM $\alpha$ cluster model  of Ref. \cite{Suzuki1976}, which reproduces 
 almost  all the energy levels  
well up  to $E_x$$\approx$13 MeV and the  electric transition probabilities   in $^{16}$O. 
The calculated $B(E2:$ $2_1^+$$\rightarrow$0$^+_1$), 7.5 $e^2$fm$^4$,  agrees
 well with the experimental data,  7.6 $e^2$fm$^4$  \cite{Okabe1995}. 
We take into account  the important transition densities 
available in Ref.\cite{Okabe1995}, i.e., g.s $\leftrightarrow$  $3^-$ (6.13 MeV) and 
$2^+$ (6.92 MeV)  in addition to all the
 diagonal densities.  
$\rho_{kl}^{\rm (^{12}C)} ({\bf r})$ represents the diagonal ($k=l$) or transition ($k\neq l$)
 nucleon density of $^{12}$C  calculated using the microscopic three $\alpha$ cluster model 
in the resonating group method \cite{Kamimura1981}. This model reproduces the structure of 
 $^{12}$C  well  and the  wave functions     have  been checked for many experimental  data,
 including charge form factors and electric transition probabilities 
\cite{Kamimura1981}. 
The calculated $B(E2:$ $2_1^+$$\rightarrow$0$^+_1$), 9.3 $e^2$fm$^4$ and 
$B(E3:$ $3^-$$\rightarrow$0$^+_1$),  124 $e^2$fm$^6$, agree well with the experimental data,
 7.8 $e^2$fm$^4$ and 107 $e^2$fm$^6$, respectively \cite{Kamimura1981}. 
In the coupled channels calculations we  take into account  
the   0$^+_1$ (0.0 MeV), $2^+$ (4.44 MeV),   and 3$^-$ (9.64 MeV) states of $^{12}$C.
 The mutual excitation channels in which both   $^{12}$C and $^{16}$O are excited simultaneously
 are not   included.  For the  effective interaction   $v_{\rm NN}$     we use  
 the DDM3Y-FR interaction \cite{Kobos1982}, which takes into account the
 finite-range    exchange effect \cite{Khoa1994}.
 An imaginary potential (non-deformed) is introduced   phenomenologically
 for all the diagonal potentials to take into
 account the effect
of absorption due to other channels, which was successful in the recent coupled channels  studies of $^{16}$O+$^{12}$C 
 rainbow scattering   \cite{Ohkubo2014,Ohkubo2014B}.  Off-diagonals are
assumed to be  real.

\par
In Fig.~1  angular distributions of elastic and inelastic $^{16}$O+$^{12}$C  scattering at 
$E_L$=170-281 MeV,   calculated  using the coupled channels method 
  including   coupling to the $2^+$ and $3^-$ states of $^{12}$C  and  $^{16}$O,
 are displayed in comparison  with  the  measured experimental data.
The same  imaginary potentials  are used for all the channels for simplicity.
 The potential parameters used and the  values of the volume integral per
 nucleon pair of the double folding (DF) potential, $J_V$,  are given in Table I.
 We found that  the  DF potential works well without introducing a normalization factor. 
  The values of the volume integral per nucleon pair   are consistent 
with those used in other
 DF optical  model  calculations \cite{Nicoli2000,Ogloblin2000,Khoa1994,Brandan2001}.
The DF potentials used
   belong to the same global potential family
 found in the $E_L$=62-124 MeV  \cite{Nicoli2000} and  
$E_L$=132-1503 MeV regions \cite{Ogloblin2000,Khoa1994}.
The  agreement with the  experimental data in elastic scattering   is comparable to the 
 optical model calculations in Refs. \cite{Ogloblin2000,Ogloblin2003}. The evolution of the
 Airy minimum in elastic scattering  is consistent with that in the  lower energy region 62-124 MeV
 \cite{Nicoli2000} and 132-260 MeV \cite{Ogloblin2000} studied with the single channel optical
 potential  model.
 The calculated  elastic scattering cross sections  are decomposed into  farside and nearside
 components. We see   that the angular distributions are dominated by
 the refractive farside scattering   in this energy region.
This means that inelastic scattering  in the energy region is dominated  by refractive waves.
In fact, we have confirmed that inelastic scattering to the $0_2^+$ state at 7.65 MeV, for which
the decomposition of the calculated scattering amplitude  into its farside and nearside components
 is  much easier because of no magnetic substates, is also dominated by farside refractive scattering.
The characteristic features of the experimental angular  distributions  in inelastic scattering
 are reproduced well by the calculations.
 We  see a  clear minimum caused by refractive inelastic scattering in the 
$\theta_{c.m.}$=95-85$^\circ$ region of
 the calculated angular distributions at 170 and  181  MeV, 
which makes it possible to identify the Airy minimum 
   at $\theta_{c.m.}$$\approx$87 and   83$^\circ$,  respectively, in the experimental 
angular distributions   in inelastic rainbow scattering.
At 200 and 260 MeV the Airy minimum is also seen at 
$\theta_{c.m.}$$\approx$75 and 53$^\circ$ in the calculated angular distributions, 
respectively, although   the corresponding minimum
is  fading in the experimental angular distributions.

\begin{table}[t]
\begin{center}
\caption{ \label{Table I}
\\
The   volume integral per nucleon pair $J_V$  of the 
DF potential  and the   imaginary potential parameters
 used in the coupled channels calculations  in Fig.~1.
}

\begin{tabular}{cccclc}
 \hline
  \hline
$E_{L}$ & $J_V$ (el)      & $W_V$  &$R_V$ &$a_V$   &  $J_V$ (inel)\\
\hline
 (MeV) & (MeV fm$^3$) &(MeV) &(fm) &(fm) &(MeV fm$^3$)    \\    \hline
  170 &   306                & 15.5   & 5.7 & 0.65        &  303           \\
 181 &   304                 &  16.5  & 5.7 & 0.55        &   301              \\
 200 &   301                 &  18.0   & 5.6 & 0.65       &   298              \\
 260 &    290               &  18.5   & 5.6 & 0.60        &   287             \\
 281 &    285                &  18.0   & 5.6 & 0.60      &   283                \\   
 \hline                          			              			          
 \hline                          				   
\end{tabular}
\end{center}
\label{Table1}
\end{table}

\begin{figure}[t]
\includegraphics[keepaspectratio,width=8.7cm] {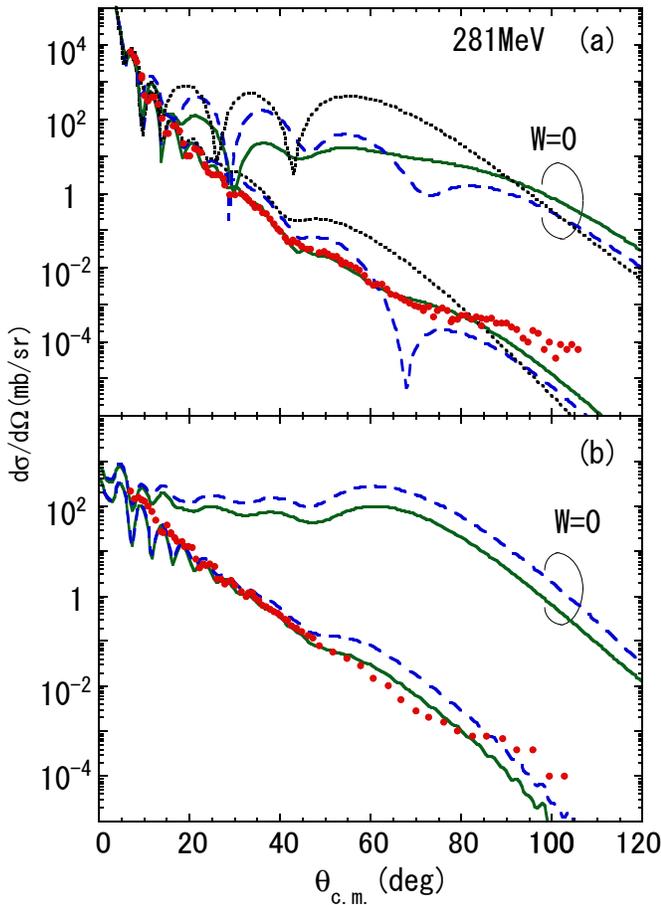}% ,angle=-90Here is how to import EPS art
 \protect\caption{\label{fig.2} {(Color online) 
 Angular distributions in $^{16}$O+$^{12}$C  (a)   elastic and (b) inelastic scattering 
to the $2^+$ state  of  $^{12}$C  at $E_L$=281 MeV  calculated using the 
coupled channels method  with  the imaginary potentials in Table I 
 %(solid lines)
 are compared with those calculated by switching off the imaginary potential
% (dotted lines) 
and the experimental data (points).
 The solid  and   dashed  lines represent the coupled channels  calculations with 
   coupling to the $2^+$ 
and $3^-$ states of $^{12}$C and $^{16}$O, and  those with
 coupling to the $2^+$  state of $^{12}$C  only, respectively.
The dotted  lines represent the  single channel calculation. 
}
 }
\end{figure}

\begin{figure}[tbh]
\includegraphics[keepaspectratio,width=8.7cm] {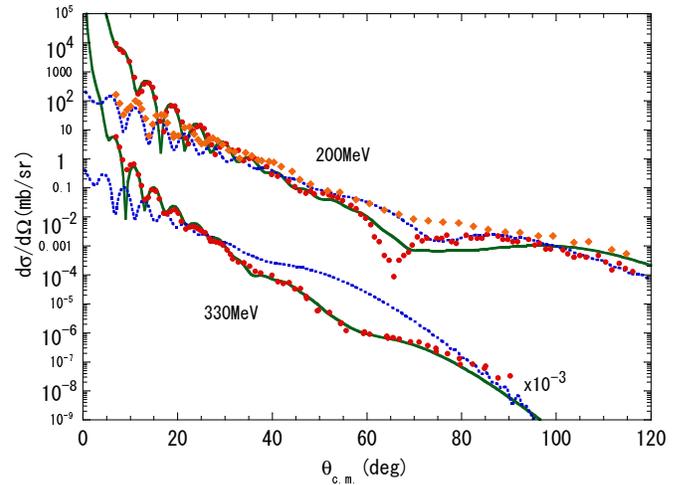}
 \protect\caption{\label{fig.3} {(Color online) 
  Angular distributions in inelastic $^{16}$O+$^{12}$C   scattering to the 2$^+$ 
state of
 $^{12}$C  (dotted line) and elastic scattering (solid line) at 200 MeV and 330 MeV calculated 
using the coupled channels method  including coupling to the $2^+$ 
and $3^-$ states of $^{12}$C and $^{16}$O are compared with the experimental inelastic (squares)  and elastic (points)
 data \cite{Demyanova}. For  330 MeV the  potential parameters in  Table I of Ref. 
\cite{Ohkubo2014} are used.
}
}
\end{figure}

\par
 In order to see  the persistence  and evolution  of the Airy minimum clearly 
at the higher energies, 
which is obscured by the imaginary potential, we show in Fig.~2  the   elastic and 
inelastic scattering angular distributions  at 281 MeV calculated with the real potential in 
Table I but 
 switching off the  imaginary potential.  Very recently it has been reported  \cite{Ohkubo2014} 
 that 
  the Airy minimum in Fig.~2(a)  at  $\theta_{c.m.}$$\approx $65$^\circ$ is caused by  the coupling to 
the $2^+$ state of $^{12}$C and that  coupling
 to the other excited states of $^{12}$C and $^{16}$O plays  a role to obscure this Airy minimum
 improving the agreement with the experimental data in the relevant 
angular region.    The role of the $2^+$ state of $^{12}$C
 is  clearly seen in Fig. 2(a)  in the calculations where the imaginary potential is switched off.
The Airy minimum caused by the coupling to the  $2^+$ state of $^{12}$C  (dashed line) 
is much  clearer.
The Airy minimum at  $\theta_{c.m.}$$\approx$40$^\circ$ in Fig.~2(a) appears even in the single channel
 calculation (dotted line). The full coupled channels calculation with 
coupling  to the $2^+$ and $3^-$ states of $^{12}$C and $^{16}$O  without the imaginary potential
(solid line) gives an Airy minimum at  $\theta_{c.m.}$$\approx$40$^\circ$, which is similar to the
 single channel calculation. Thus 
the Airy minimum at $\theta_{c.m.}$$\approx$40$^\circ$ is an ordinary
   nuclear rainbow caused by the refractive DF potential,
   and the Airy minimum at the larger angle 
  $\theta_{c.m.}$$\approx$65$^\circ$ in Fig.~2(a)
was claimed to be a new kind of  Airy minimum of a secondary rainbow  dynamically caused predominantly by 
 coupling to the deformed   $2^+$ state of $^{12}$C \cite{Ohkubo2014}.
 The role of  the excitation to the 2$^+$ state of $^{12}$C  was  also investigated   at other energies 
using the potential parameters in Table I.
 On the other hand, as for the inelastic scattering to the $2^+$ state of $^{12}$C, the calculations
 without the imaginary potential  in Fig.~2(b) show an Airy minimum at  $\theta_{c.m.}$$\approx$47$^\circ$.
 Thus the small minimum at   $\theta_{c.m.}$$\approx$47$^\circ$   in the experimental angular
 distribution, which is  close to the Airy minimum in elastic scattering,
 is assigned to be  an Airy minimum in inelastic scattering. 
The Airy minimum in inelastic scattering at 260, 200, 180 and 170 MeV can be similarly 
located  at $\theta_{c.m.}$$\approx$ 53, 75,
 83 and 87$^\circ$, respectively.  
 The broad   Airy maximum  at 281 MeV seen clearly in Fig.~2(b) in the calculations    without the
 imaginary potentials, centered at $\theta_{c.m.}$$\approx$60$^\circ$,
  seems to have some  effect 
 on the elastic  scattering. 
\begin{figure}[b]
\includegraphics[keepaspectratio,width=8.7cm] {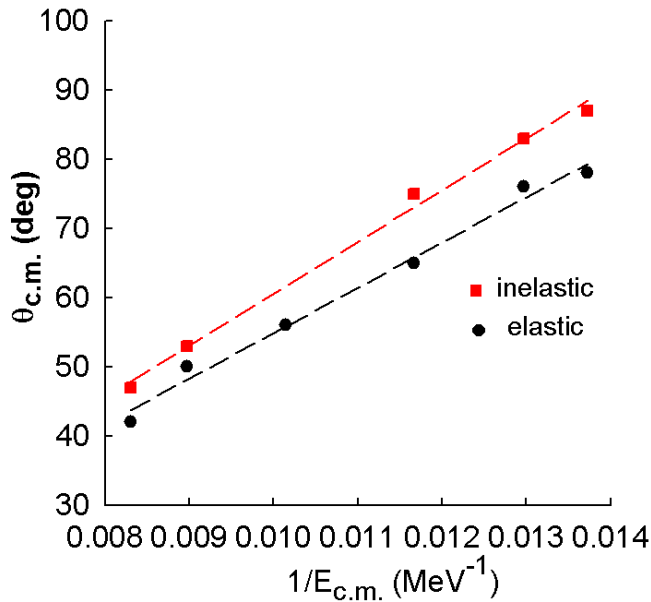}% ,angle=-90Here is how to import EPS art
 \protect\caption{\label{fig.4} {(Color online) 
 The positions of the Airy minimum  observed  in  inelastic scattering to 
the $2^+$ state of
 $^{12}$C  (filled square) and elastic scattering (filled circle)  of 
   $^{16}$O+$^{12}$C are displayed as a function of the inverse c.m. energies.
The lines are drawn by fitting the data.}
}
\end{figure}

\begin{figure}[bht]
\includegraphics[keepaspectratio,width=8.7cm] {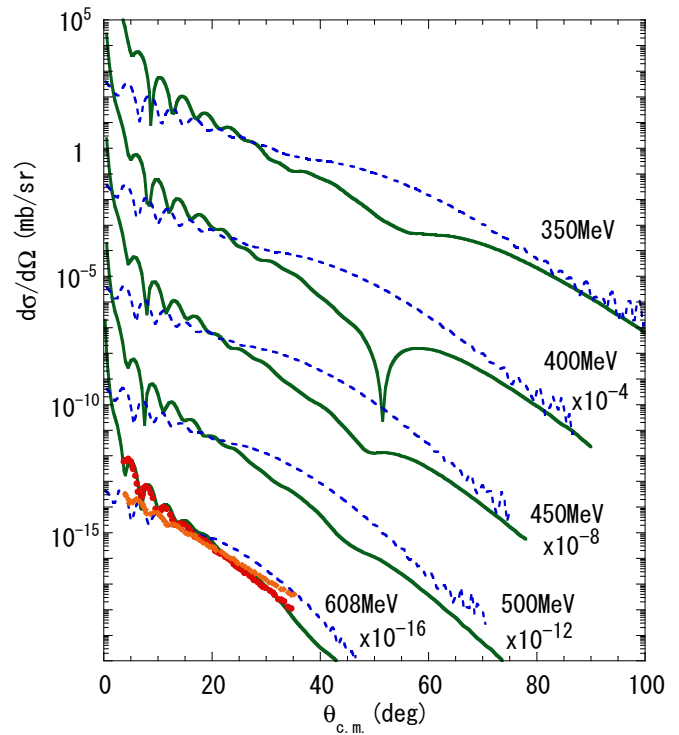}% ,angle=-90Here is how to import EPS art
 \protect\caption{\label{fig.5} {(Color online) 
  Angular distributions in  $^{16}$O+$^{12}$C elastic  (solid line) and inelastic 
scattering to the $2^+$ state of  $^{12}$C   (dashed line)  at $E_L$=350,  400, 450,   500  and 608 MeV  
 calculated using the coupled channels  method including   coupling to the $2^+$ 
and $3^-$ states of $^{12}$C and $^{16}$O.
 The experimental data  (points) at  608 MeV are taken from Ref. \cite{Brandan1986}.
}
}
\end{figure}

\par
 To show that the inelastic scattering to  the $2^+$ state of $^{12}$C  is more enhanced
than the elastic scattering, the experimental and calculated angular distributions at
 200 and 330 MeV are compared  in Fig.~3.
 At 200 MeV we see that the observed  cross sections in  inelastic  scattering are   larger than  
those in elastic scattering in the intermediate angular region.
 Also,  at 330 MeV the calculated inelastic scattering is greatly enhanced compared with the 
observed elastic scattering cross sections at intermediate angles.
  We note that the broad peak of the Airy maximum in inelastic scattering appears 
in the angular region  where the   Airy minimum appears  at     $\theta_{c.m.}\approx 60^\circ$
 in elastic scattering.
The coupling between the elastic scattering and inelastic scattering has a large effect on the 
 behavior of the angular distributions in  the relevant angular region and causes a new kind of
 dynamically induced Airy minimum in elastic scattering \cite{Ohkubo2014}.

\par
 In Fig.~4 the energy evolution of the angles of the Airy minimum in elastic scattering and inelastic
 scattering is displayed as a function of inverse c.m. 
energies.
We see that the Airy minimum in inelastic scattering is slightly shifted toward the larger angles.
This shift is  caused by
the excitation energy effect as discussed in Ref.  \cite{Michel2004A}, because the size of the
 $2^+$ state of $^{12}$C is almost as large as that of the band 
head ground state $0^+$. This  is confirmed by the   
    volume integrals in Table  I, which are almost the same  for elastic and 
  inelastic scattering. The positions of the Airy minima in inelastic scattering
are  approximately on the straight line similar to those in elastic
 scattering \cite{Ogloblin2003}, which gives  a support to the present 
assignment  of the Airy minimum in inelastic scattering. 

\par
  In Fig.~5 we show  the calculated angular distributions in  elastic and inelastic 
$^{16}$O+$^{12}$C  scattering at $E_L$=350-608 MeV. The  imaginary potentials
 were interpolated from Table I in Ref. \cite{Ohkubo2014}.   The calculations predict
 an Airy  minimum of the secondary  bow in  elastic  scattering caused by coupling to 
the inelastic channels  at the higher energies such as 
 a clear Airy minimum at  400 MeV.  On the other hand, in  inelastic scattering
the  emergence of the  Airy minimum seems obscured because of  absorption.
 However,  the rainbow pattern with
 the Airy maximum  accompanying the  fall-off of the cross sections toward the large angles,
 which was  discussed for the $^{16}$O+$^{16}$O system  by Khoa {\it et al.} \cite{Khoa2005}, 
persists.
 Because the dynamical Airy minimum in elastic scattering    is mostly   brought about by coupling
 to the $2^+$ state  of $^{12}$C \cite{Ohkubo2014}, the  appearance of this Airy minimum, for example at
 400 MeV,  seems to be related to  the  persistence of the nuclear rainbow in inelastic scattering 
to the  $2^+$ state.
At  608 MeV,  the experimental angular distribution  of  Brandan {\it et al.}
 \cite{Brandan1986} shows the fall-off of the cross sections  both in  elastic and inelastic scattering 
 and no Airy minimum is confirmed.
 At 608 MeV,  a   dynamical Airy minimum of the secondary bow in elastic scattering is no longer
 perceptible,   even in the calculations  with $W_V=0$, suggesting that the coupling to the 2$^+$ state is
not   of specific  importance in this high energy region.

%added
Finally we discuss the reasons of the   obscurity of  the experimental Airy minimum in the inelastic
 scattering data compared with calculated results.  The contribution of the
     $\alpha$ particle transfer is not important for the obscurity of the Airy minimum. 
In fact,    the coupled  reaction channels calculations of  $^{16}$O+$^{12}$C  scattering  
in Ref.  \cite{Rudchik2010}     showed that  it is more than three orders of magnitude
 smaller than the experimental data in the relevant angular region.
 We note that   the  present  calculations  take into account the one nucleon exchange effect, 
which is suggested to prevail over other transfer reactions \cite{Rudchik2010},  by using 
effective  interaction DDM3Y, in which   the knock-on exchange   effect is incorporated 
\cite{Kobos1982,Brandan1997}. 
As for the coupling to the other  excited states of $^{12}$C such as the $0^+_2$,  $0^+_3$, 
 and    $2^+_2$  states, we have confirmed in the extended coupled channels calculations that its 
contribution to the Airy minimum in inelastic scattering is not significant. 
In the present 
calculations mutual excitations, in   which both the $2^+$ of $^{12}$C and the excited states of
 $^{16}$O are  excited simultaneously, are not included.
 To take into account  
the flow of the flux via the $2^+$ state of $^{12}$C to the excited $2^+$ and
 $3^-$ states of $^{16}$O phenomenologically, 
 in Fig.~1   angular distributions calculated by using a slightly stronger imaginary potential 
($W_V=24$, $R_V$=5.0 and $a_V$=0.8)  for the channels
 are  displayed at 200 MeV by the blue 
%thin solid
 dotted lines. We see that  the calculated  Airy minimum 
 preceding the first Airy maximum (the broad rainbow shoulder) is obscured
 significantly in close to the experimental data by this imaginary potential effect. 
Because a phenomenological imaginary potential cannot correctly  replace a microscopic treatment of
 coupling  as was discussed for a secondary bow in Ref. \cite{Ohkubo2014} and ripples in the
 nuclear rainbow in Ref. \cite{Ohkubo2014B},  an accurate microscopic treatment of these channels
 coupling may be necessary to investigate the further obscurity of the Airy minimum of the 
experimental data.

\section{SUMMARY}
 \par
%To summarize,  
Inelastic $^{16}$O + $^{12}$C rainbow scattering to 
the  $2^+$  (4.44 MeV)  state of $^{12}$C  was measured at $E_L$ = 170-281 MeV. 
A systematic analysis of the existence  and evolution 
of the Airy minimum   in the angular distributions in   inelastic  rainbow scattering
 was done using the coupled channels method 
  with an extended double folding  potential 
derived from the realistic  wave functions for $^{12}$C  and $^{16}$O calculated with a 
 microscopic  $\alpha$ cluster model with  a finite-range  density-dependent 
  nucleon-nucleon force. 
The coupled channels  analysis of the measured inelastic $^{16}$O +$^{12}$C  scattering data
 shows consistently some Airy-like
structure in the inelastic scattering cross sections for the first $2^+$
state of $^{12}$C, which is somewhat obscured and still not clearly visible
in the measured data. 
The Airy minimum was identified from the analysis and it was found that 
the Airy minimum in inelastic scattering is   shifted
  backward compared with 
that in elastic scattering.
 The existence of  rainbows in  inelastic scattering seems
 to be  responsible for creating  a dynamical Airy minimum of the secondary
 rainbow in elastic scattering.
It is intriguing to study the Airy minimum in  inelastic heavy ion rainbow  scattering
 experimentally and theoretically  and to reveal the relationship between 
 the dynamically induced  secondary bow in elastic scattering and the structure  of the involved
 nuclei.

The work was partly supported by Russian Scientific Foundation 
(Grant No. RNF14-12-00079).
 Two of the authors (S.O. and Y.H.) would like to thank the Yukawa Institute
 for Theoretical Physics for
 the hospitality extended  during a stay in   2014. 
Part of this work was supported by the Grant-in-Aid for the Global
 COE Program ``The Next
 Generation of Physics, Spun from Universality and Emergence'' from the Ministry 
of Education, Culture, Sports, Science and Technology (MEXT) of Japan.

\end{document}